\documentclass[10pt,final,conference]{IEEEtran}
\ifCLASSINFOpdf
\else
\fi
\ifCLASSINFOpdf
  \usepackage[pdftex]{graphicx}
  \graphicspath{{../pdf/}{../jpeg/}}
   \DeclareGraphicsExtensions{.pdf,.jpeg,.png}
\else
  \usepackage[dvips]{graphicx}
   \graphicspath{{../eps/}}
   \DeclareGraphicsExtensions{.eps}
\fi
\usepackage{subfigure}
\usepackage{xcolor}
\usepackage{listings}

\IEEEoverridecommandlockouts

\begin{document}
%
\title{ABMQ: An Agent-Based Modeler and Simulator for Self-Organization in MANETs using Qt\thanks{This work was supported by the Finnish Funding Agency for Technology and Innovation (Tekes) under the Celtic-Plus ``Green Terminals for Next Generation Wireless Systems" (Green-T CP8-006) project.}}

\author{
    \IEEEauthorblockN{Mohammad Noormohammadpour$^{1}$, Mohammad Javad Salehi$^{1}$, Seyed Mohammad Asghari Pari$^{1}$\\Babak Hossein Khalaj$^{1}$, Hamidreza Bagheri$^{2}$, Marcos Katz$^{2}$}
    \IEEEauthorblockA{$^{1}$Department of Electrical Engineering, Sharif University of Technology, Tehran, Iran\\$^{2}$Center for Wireless Communications, University of Oulu, Finland, 90014\\\{noormohammadpour, mjsalehi, asgharipari\}@ee.sharif.ir, khalaj@sharif.ir, \{hamidreza.bagheri, marcos.katz\}@ee.oulu.fi}
}
\maketitle

\begin{abstract}
Agent-Based Modeling and Simulation (ABMS) is a simple and yet powerful method for simulation of interactions among individual agents. Using ABMS, different phenomena can be modeled and simulated without spending additional time on unnecessary complexities. Although ABMS is well-matured in many different fields such as economic, social, and natural phenomena, it has not received much attention in the context of mobile ad-hoc networks (MANETs). In this paper, we present ABMQ, a powerful Agent-Based platform suitable for modeling and simulation of self-organization in wireless networks, and particularly MANETs. By utilizing the unique potentials of Qt Application Framework, ABMQ provides the ability to easily model and simulate self-organizing algorithms, and then reuse the codes and models developed during simulation process for building real third-party applications for several desktop and mobile platforms, which substantially decreases the development time and cost, and prevents probable bugs that can happen as a result of rewriting codes.
\end{abstract}


\section{Introduction}

A mobile ad-hoc network (MANET) is comprised of several mobile devices connected by means of short-range wireless links, in which each device is a router for other devices as well as being an end host. In MANETs, each device is free to operate itself independently, which provides a very flexible environment for users. Furthermore, they do not require a pre-existing infrastructure, and they use multi-hop routing protocols for communication [1]. 

Despite all clear advantages and features of MANETs, there are also many challenging issues which have limited their usability. Some of these challenges are dynamic network topology, limited resources available in participant devices, security issues [2], and selfish behavior of devices causing problems such as denial of service (DOS) [3]. Such issues are more disruptive as network grows in size and becomes more complex.

Self-Organization in MANETs provides efficient solutions for building large distributed systems, in which participant hosts communicate with one another in order to manage the network without the guidance of an external source. In general, self-organized communication networks are reliable, flexible, robust, and cost-effective [4].

As presented in [4], self-organization can be implemented in two different levels: infrastructure and application. While the first one aims at self-organization in various communication layers of network management (Data Link, Network, Transport), the second one intends to provide self-organization above such layers. In either level, a proper self-organizing algorithm is required in network. 

Obviously, such algorithms should be evaluated under many different circumstances in order to prove their feasibility. Subsequent to each test, the algorithm is modified, probable bugs and errors are corrected, and another test is performed. This time-consuming process continues until all tests are successful and the required performance measures are met.

Similar to other simulation platforms in the network context [5] [6] [7], in this paper, we present a new simulation platform, called ABMQ, and based on Qt\footnote{Qt Application Framework. [Online]. Available: http://www.qt-project.org}, which is suitable for modeling and simulation of self-organization in complex networks, and particularly MANETs. The main focus of ABMQ is on simulation of self-organization at application level while preserving the ability for simulating the infrastructure level algorithms. Furthermore, it provides many useful features for reusing the codes and models developed over the simulation process in order to build real third-party applications.

An Agent-Based approach is taken in implementation of ABMQ. In Agent-Based Modeling and Simulation (ABMS), autonomous individuals (i.e., agents) communicate and interact with each other, and also the environment, while their effect on the system is assessed as a whole [8]. In ABMQ, each participant device is considered an agent. Such definition fits devices making up a MANET, where autonomy is the most important characteristic of participants. 

Following up the ABMS, several simulation platforms have been proposed in this context [9]. Swarm is the first platform, development of which started in Objective-C, then followed by Java and named as Java Swarm [10]. Repast is another Java-Based platform which is mainly designed for modeling and simulation in social sciences [11]. MASON is also a Java-Based multi-agent toolkit in which the speed of computation is considered the priority [12]. Among all these simulators, NetLogo is the most professional platform which is also based on Java [13]. Using an initiative high-level programming language, NetLogo is mainly designed for simulation of natural and social phenomena.

However, none of these platforms provides a dedicated environment for modeling and simulation of self-organizing algorithms in MANETs, while enabling the capability to reuse the created models for developing real applications. The process of reusing the implemented models and codes can significantly decrease the required time for developing such applications. Since such algorithms can target smartphones, tablets, laptops, and many other portable devices with different operating systems (Android, iOS, Symbian etc.), using a cross-platform development environment can substantially facilitate the deployment process.

Although it was possible to develop ABMQ upon programming environments such as Java Platform, we decided to base it on Qt. The main reason for this selection was a number of key properties of Qt, such as native code support for target operating systems, and its optimized speed due to C++ infrastructure.

The rest of this paper is organized as follows: ABMQ is proposed in section II. Section III presents two use cases of ABMQ, and finally, section IV concludes the paper.


\section{The Simulation Platform}

In this section, we introduce ABMQ. First, we briefly introduce our development environment, the Qt Application Framework. The architecture of ABMQ is presented next.

\subsection{The Qt Application Framework}

As stated earlier, our main objective is to simulate self-organizing algorithms in application level, mainly targeting portable and programmable devices such as smartphones and tablets. As these devices utilize a number of different operating systems, it is efficient to use programming frameworks with the capability of cross-platform development.

One such cross-platform programming framework is the Qt Application Framework, which has enabled programmers to write a single piece of code and deploy it on many operating systems. By means of this feature, users can easily use the codes they have written for simulation purposes to develop real applications on smartphones, tablets, and laptop computers. Currently, using Qt, programmers are able to develop programs for a variety of desktop and mobile operating systems, including Microsoft Windows, Mac OS X, Linux, BlackBerry, Symbian, MeeGo, Windows CE, Maemo; and also support for Android and iOS is to follow soon.

Furthermore, as Qt uses C++ programming language with native code support for target operating systems, applications developed on this platform have very good performance on target devices. Qt is also available under open source licenses such as LGPL.

\subsection{ABMQ Architecture}

Figure 1 represents the architecture of ABMQ. This architecture consists of three major parts: GUI, Agents, and Simulation Management Block (SMB). The connection among these three main parts is established using Qt's advanced event dispatching method called Signals and Slots, as well as some global variables. In the following, we describe the tasks of each part:

\begin{figure}[h]
\begin{center}
\includegraphics[width=85mm]{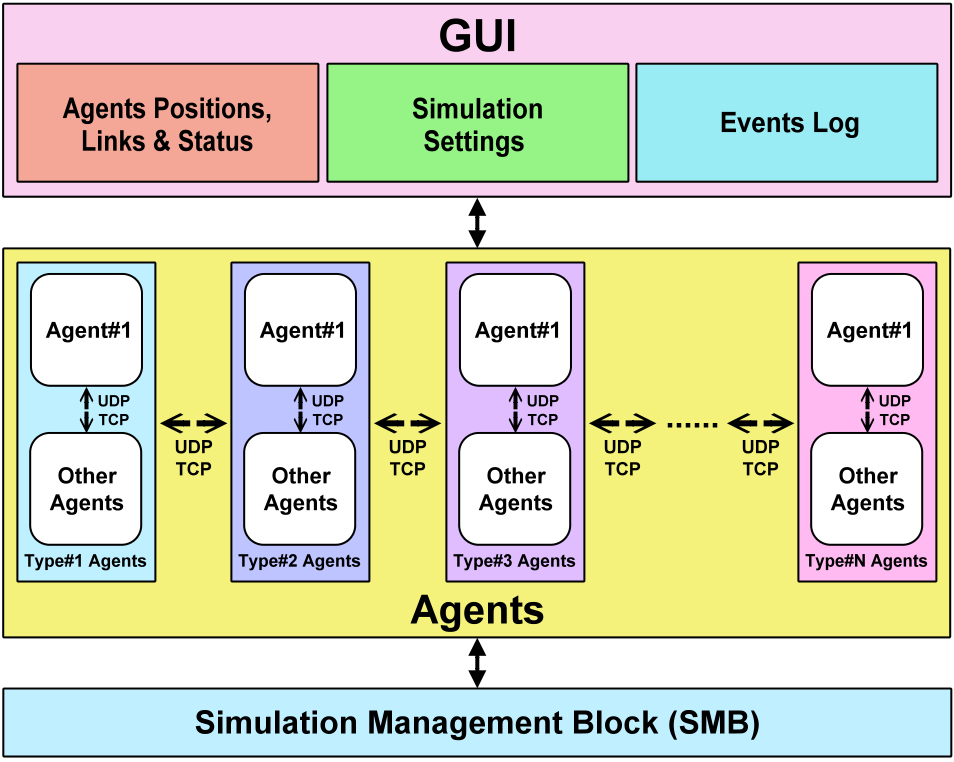}
\caption{ABMQ Architecture}
\end{center}
\end{figure}

\textbf{GUI (Graphic User Interface):} The GUI illustrates all necessary and important information for users, and also enables them to select desired settings for the simulation process. GUI section can be changed depending on the algorithms implemented for simulation. As presented in Figure 1, three major parts constitute the GUI, each of which providing a different task:

\begin{enumerate}
\item The first and most important part is the Simulation Settings window. This window contains all information about agents, including their exact location, identification number, and movement pattern. In addition, the required settings for simulation process can be applied from this window.

\item The other major part of GUI is the graphical view of all agents, including (but not limited to) their location, identification number, and connection links among them.

\item The last major part is the Events Log window. The purpose of this window is to provide an instant view of all important interactions among agents.
\end{enumerate}

\textbf{Agents:} Agents constitute the core of our simulation platform. Each agent simulates the behavior of one real participant device (for example, a smartphone, a tablet, or a laptop) in MANETs. In ABMQ, each agent is an object created based on a C++ class. Users should define all of the required protocols and algorithms in this class. Since agents should be independent from each other (because in a real world they already are), in ABMQ each agent is created inside a separate thread. This process is done through inheriting such C++ classes from QThread, the base class for platform-independent thread management in Qt. As threads are managed independently by the host operating system, each one of the agents is independent from the others.

As shown in Figure 1, different types of agents are supported in ABMQ. Agents from the same type are objects created from an identical C++ class, while agents from different types are constructed from different C++ classes. Figure 2 represents a simple class from which agents can be created. For example, two different classes can be defined, one for smartphone agents and another for tablet agents. Then, for simulation of interactions among three smartphones and two tablets, three objects should be created from the first C++ class and two objects form the second one.

\begin{figure}[h]
\begin{lstlisting}
#include <QThread>
#include <QUdpSocket>
#include <QTimer>

class SampleAgent : public QThread
{
	Q_OBJECT
public:
	QUdpSocket^* udpmodule; // communication module
	QTimer^* timer; // timer for management procedures
	int X,Y; // location variables
	// other variables
	SampleAgent(int init_X, int init_Y, /* other initialization inputs such as UID, etc. */)
	{
		timer = new QTimer;
		connect(timer,SIGNAL(timeout()),
			this,SLOT(manage()));
		timer->setInterval(/* timer interval */);
		
		udpmodule = new QUdpSocket(this);
		connect(udpmodule,SIGNAL(readyRead()),
			this,SLOT(DataReceived()));
		udpmodule->bind(QHostAddress::LocalHost,/* listen port */);
		
		X = init_X;
		Y = init_Y;
		
		// other initialization codes
	}
	// other functions
	
public slots:
	void manage() // called on timer timeouts
	{
		// add management codes here
	}
	void DataReceived() // called when a packet is received
	{
		// add message processing codes here
	}
};
\end{lstlisting}
\caption{A sample class for agents}
\end{figure}
 
When users choose to add new nodes to the simulation environment, the GUI part, which is interacting with the user, creates new objects based on the selected agent type and configures them with the settings selected by users. Figure 3 represents a sample code in which 50 agents are created from the class which was previously defined for agents in Figure 2. Furthermore, a unique identification number (UID) is assigned to each agent. In this way, hundreds of autonomous agents can be created with different types and configurations, each of which running in a separate thread and the behavior of each agent can be monitored precisely.

\begin{figure}[h]
\begin{lstlisting}
// somewhere in the main thread (GUI thread)
// create 50 agents from the SampleAgent class
SampleAgent^* agent;
int X = 0,Y = 0;
for(int i=0;i<50;i++)
{
	agent = new SampleAgent(X, Y, /* other initialization inputs such as UID, etc. */);
	agent->start();
	X += 10; // change X for the next agent
	Y += 5; // change Y for the next agent
}
\end{lstlisting}
\caption{Creating agents from the predefined class}
\end{figure}

While it is possible to use mere simulation models for establishing communication among the agents, we decided to use standard network protocols for this purpose. Using this implementation method enables users to turn their written codes into real applications much faster and easier. Furthermore, it also helps in generating more realistic results. The communication protocol can be either TCP or UDP (or other protocols that the underlying operating system supports). While UDP is known for simplicity and speed, TCP provides reliability, in-order packet transfer, flow control, and congestion control. In order to use different communication protocols in ABMQ, users can create an object from the base class of their required protocol and then configure it with the appropriate parameters.

As an example, for implementing UDP module in agents, users can create an object from QUdpSocket class and then configure its listen IP address and port number for receiving the UDP datagrams from other agents. In addition, users can use the same object for sending UDP datagrams to other agents by means of target agent's IP address and port number. For the purpose of simulation, the listen IP address for all agents is set to one of the loop back IP addresses (e.g., 127.0.0.1). However, each agent requires a unique listen port number for communication. Such port number can be assigned to agents based on their UIDs. As an example, the sample agent presented in Figure 2 uses UDP as its communication protocol.

\begin{figure}[h]
\begin{center}
\includegraphics[width=85mm]{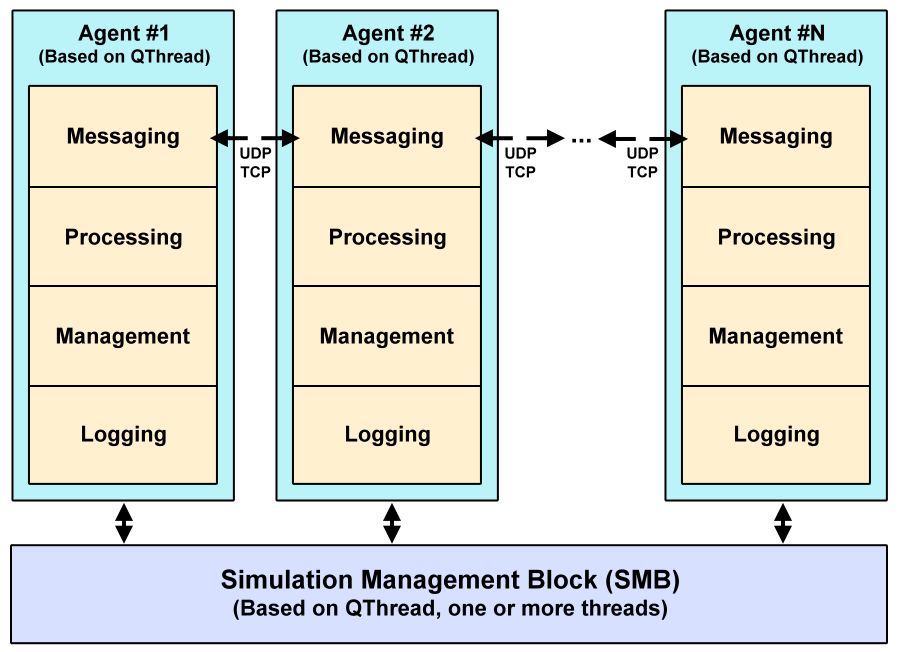}
\caption{Agents Functional Architecture}
\end{center}
\end{figure}

Figure 4 represents the functional architecture of agents. As can be seen, each agent consists of four different functional components:

\begin{enumerate}
\item \textbf{Messaging:} This component enables agents to communicate with each other. As mentioned earlier, the communication protocol can be selected based on the requirements.

\item \textbf{Processing:} The purpose of this component is to process different tasks of an agent. An example of such tasks is processing of incoming information and making proper decisions and responses based on it.

\item \textbf{Management:} Including a wide variety of different tasks, this is the most important component in an agent. Some of the tasks performed by this component are movement of the agent, discovery of neighbor agents, and updating the status of the agent in GUI.

\item \textbf{Logging:} The important status information, variables, and parameters regarding each agent are logged continuously. Using this information, the required simulation results can be obtained.
\end{enumerate}

\textbf{Simulation Management Block (SMB):} In addition to GUI and agents, a management mechanism is also required among the agents. Such management does not question the independent nature of agents; besides, it satisfies some low-level or environment related requirements of agents for the simulation process. Synchronization among agents (if required) and multi-hop routing in application level simulations are two examples of tasks performed by SMB. As can be seen in Figure 4, SMB is connected to all agents. In fact, SMB plays a complementary role for management component inside each agent. In addition, since SMB is connected to all agents, it can be helpful for extracting some of the simulation results. 

Based on the required tasks, SMB can be very complex or very simple (it can even be unnecessary). SMB is comprised of one or more separate threads while each thread performs one of the required tasks. Since such tasks are completely reliant on the simulation requirements, it is not possible to provide a functional architecture for this section.

\section{Simulation Use Cases}

In this section, we present two different simulation use cases for ABMQ. The first case is a method of clustering for MANETs which incorporates a self-organizing approach [15]. In the second case, a leader selection algorithm for Mobile Clouds [16] is implemented and simulated. In addition, for the second case, we further use the implemented algorithms and models for creating a real application which can be installed on several different mobile and desktop operating systems.

\subsection{MANET Clustering with a Self-Organizing Approach}

Although MANETs provide us with many useful features, their efficiency and performance decreases as they grow in size. Partitioning a large MANET into smaller virtual networks (clusters) can considerably increase the overall efficiency in terms of power consumption, bandwidth usage, and management, which provides a better performance in return [14].

In this section, we simulate a method of clustering for MANETs, which is presented in [15]. According to this clustering method, each participant device can be a cluster head, a cluster gateway, or a cluster member. The size of each cluster is limited based on the maximum allowed number of hops between each cluster member and the cluster head, which is defined as a parameter named k.

The GUI section for this simulation case is presented in Figure 5. The three main components of GUI (settings window, graphical representation of agents, and log window) can be seen in this figure. Additional required controls, such as the option for configuring the parameter k, are added to the settings window using Qt's tool for designing GUIs, the Qt Designer.

\begin{figure}[h]
\begin{center}
\includegraphics[width=85mm]{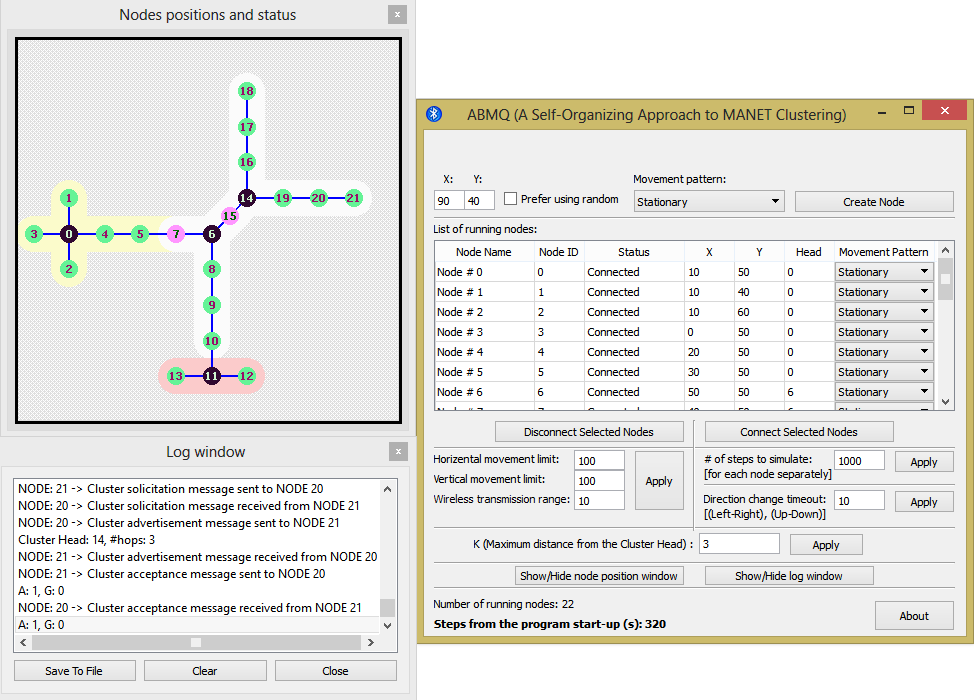}
\caption{The GUI for simulation case 1}
\end{center}
\end{figure}

Figure 6 represents a clustering example for an arbitrary network (k=3). As can be seen, different background colors are assigned to different clusters. When a device boots up, it sends cluster solicitation messages to its immediate neighbors. If it does not receive any responses, or it is unable to reach any cluster heads in equal or less than k hops, it assigns itself as a cluster head (black nodes). Conversely, if there is exactly one cluster head in k hops at most, the new device assigns itself as a member (green nodes). Moreover, devices that reach two or more cluster heads in equal or less than k hops are assigned as gateways (pink nodes). 

\begin{figure}[h]
\begin{center}
\includegraphics[width=55mm]{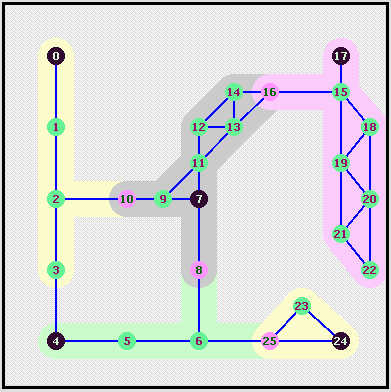}
\caption{A clustered network example}
\end{center}
\end{figure}

In this simulation case, agents use UDP protocol for communication. Each agent has a UDP module for receiving messages, which is bound to IP address 127.0.0.1 and a unique port number based on its UID. Therefore, for sending messages to any agent, it is sufficient to know its UID.

Finally, SMB also plays an important role in this simulation case. The purpose of SMB is to simulate the behavior of Network Layer by providing the routing functionality. Each agent communicates with SMB for acquiring or updating the routing information. Therefore, each agent can only communicate with agents that the implemented routing protocol authorizes it to.

\subsection{A Leader Selection Algorithm for Mobile Clouds}

The purpose of Mobile Clouds is to provide a cooperative arrangement among the communication-enabled portable devices such as smartphones, tablets or laptop computers. However, in order to make such cooperation feasible, some management method is required in Mobile Clouds. Such management method is responsible for collecting and processing status data from various devices present in the cloud, in exchange for finding resource sharing opportunities [16].

In this section, we briefly explain how we managed to implement and simulate the algorithms regarding a new framework for Mobile Cloud management as presented in [16]. According to this management framework, each participant device in Mobile Clouds can be either a leader or a client. As a result of cooperation among all devices, the one with more available resources is selected as leader. This leader then performs all management operations in the cloud, such as finding appropriate devices for resource sharing.

Figure 7 shows the GUI section for this simulation case. As can be seen, in comparison with the previous simulation case, the main components of GUI are intact, while some input parameters are changed. Figure 8 shows a simulation example using an arbitrary network. The black node is the leader agent, while green nodes are client ones. 

\begin{figure}[h]
\begin{center}
\includegraphics[width=85mm]{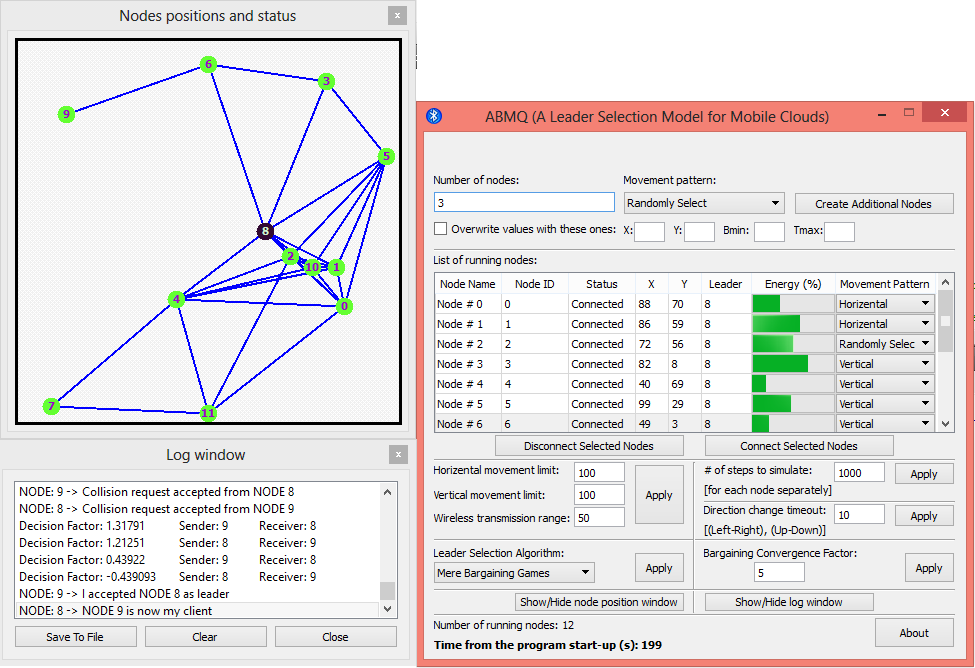}
\caption{The GUI section for simulation case 2}
\end{center}
\end{figure}

\begin{figure}[h]
\begin{center}
\includegraphics[width=55mm]{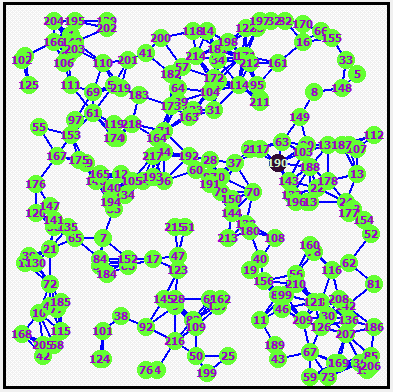}
\caption{A simulation example for Mobile Clouds}
\end{center}
\end{figure}

Similar to previous simulation example, in this case, the UDP protocol is used for communication among agents. In addition, SMB also serves similar functionality. However, the required algorithms for simulation of this case have replaced the ones of the earlier simulation example.

As mentioned before, one of the most important features of ABMQ is that users can easily convert their simulated code into real third-party applications. As an example, here we review the steps required to build one such application using the leader selection algorithm simulation codes. 

The first step for creating such third-party application is to design a proper GUI, which can be simply done by using Qt Designer, the Qt's tool for designing GUIs. 

The next step is to implement the required algorithms and functions. As mentioned earlier, all agents of the same type are objects created from a unique C++ class (agent's class), which also includes the codes users have written for testing and simulation purposes. Therefore, by including such classes in our third-party application, it is possible to reuse the previously written codes for implementing the same functionalities, which can significantly facilitate the implementation process.

Finally, the last step is to implement the communication ability in our third-party application. Since the agent's class, which was included in our application in previous step, also contains a communication module (the module which was earlier used for communication among simulation agents), such module can be again used here by means of proper configurations for IP and port settings. 

In our third-party application, the receive IP address for UDP module is set to the one of interfaces that devices are using for communication in MANET. Similarly, for sending UDP datagrams, the interface IP address of target device is used. Moreover, all devices are configured to use the same port number for receiving and sending UDP datagrams. In this way, using slight changes in the base class of agents, such classes can be reused for implementing the same functions in real applications.

Figure 9 shows our third-party application running on Symbian operating system. Using the cross-platform feature of Qt, this application can be simply compiled for other platforms, such as Microsoft Windows and Ubuntu Linux. 

As can be seen in Figure 9, the application is made up of two separate panes. The first pane provides a list of devices present in the cloud (devices running our application), including their names, IP addresses, and MAC addresses. The second pane shows some information regarding the cloud status, such as the name of the leader device and the count of members.

\section{Conclusion and Future Works}

In this paper, we presented a new simulation platform, called ABMQ, which is a powerful tool for modeling and simulation of self-organization in wireless networks and particularly MANETs. It provides two important key advantages:
\begin{itemize}
\item
It enables users to easily use the codes they have written for simulation and evaluation purposes for developing real third-party applications, especially for mobile devices. By using this feature, the required time and cost for development significantly decrease, and probable bugs happening due to rewrite of codes can be completely eliminated. 
\item
ABMQ is Agent-Based, meaning that users can simulate and evaluate their models and algorithms without requiring to spend time on unnecessary details. This makes ABMQ a very flexible, easy to use, and efficient environment for simulating interactions among mobile devices.
\end{itemize}

Although ABMQ provides a number of useful advantages, it is in the early stages of development and there are still many other subjects that are required to be addressed in order to build a better simulation environment. As an example, creating a library of tools which includes a set of predefined models that can be configured for use in different simulation scenarios can be a new subject for future researches.

\begin{figure}[h]
\begin{center}
\subfigure[Pane 1: The list of devices in the local network]{
\includegraphics[width=70mm]{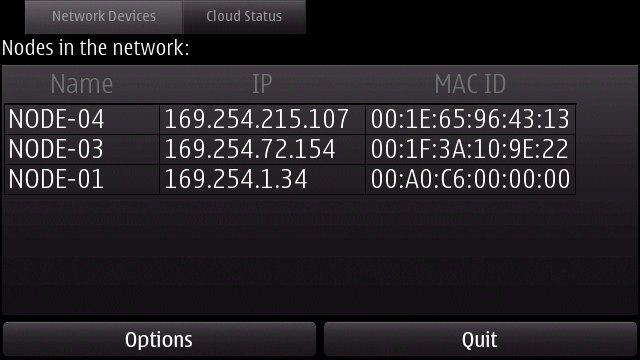}
}
\subfigure[Pane 2: Cloud status information]{
\includegraphics[width=70mm]{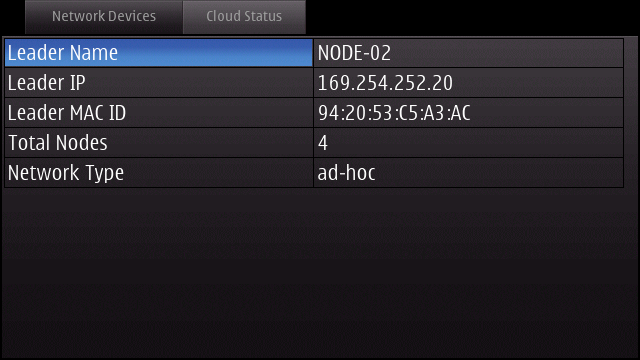}
}
\caption{Third-party application running on Symbian OS}
\end{center}
\end{figure}

\end{document}